\documentclass[aps,prl,longbibliography,
,groupedaddress, twocolumn,notitlepage]{revtex4-1}

\usepackage{amsmath,amssymb}
\usepackage{natbib}
\usepackage{graphicx}

\newcommand{\be}{\begin{equation}}
\newcommand{\ee}{\end{equation}}
\newcommand{\bse}{\begin{subequations}}
\newcommand{\ese}{\end{subequations}}
\newcommand{\ba}{\begin{eqnarray}}
\newcommand{\ea}{\end{eqnarray}}
\newcommand{\bea}{\begin{eqnarray}}
\newcommand{\eea}{\end{eqnarray}}

\usepackage{color}
\usepackage[normalem]{ulem}  

\begin{document}


\title{Linear-in-$T$ resistivity from semiholographic non-Fermi liquid models}


\author{Beno\^{i}t Dou\c{c}ot}
\email[]{doucot@lpthe.jussieu.fr}
\affiliation{Laboratoire  de  Physique  Th\'{e}orique  et  Hautes  Energies,Sorbonne Universit\'{e}s and CNRS UMR 7589, 4 place Jussieu, 75252 Paris Cedex 05, France}

\author{Ayan Mukhopadhyay}
\email[]{ayan@physics.iitm.ac.in}
\affiliation{Center for Quantum Information Theory of Matter and Spacetime, Department of Physics, Indian Institute of Technology Madras, Chennai 600036, India}

\author{Giuseppe Policastro}
\email[]{policast@lpt.ens.fr}
\affiliation{Laboratoire  de  Physique  de  l'Ecole  Normale  Sup\'{e}rieure,  CNRS,  Universit\'{e} PSL, Sorbonne  Universit\'{e}s,  Universit\'{e}  Pierre  et  Marie  Curie,  24  rue  Lhomond,  75005  Paris,  France}
\author{Sutapa Samanta}
\email[]{psss2238@iacs.res.in}
\affiliation{School of Physical Sciences, Indian Association for the Cultivation of Science, Jadavpur, Kolkata 700032, India}
\date{\today}

\begin{abstract}
We construct a semiholographic effective theory in which the electron of a two-dimensional band hybridizes with a fermionic operator of a critical holographic sector, while also interacting with other bands that preserve quasiparticle characteristics. Besides the scaling dimension $\nu$ of the fermionic operator in the holographic sector, the effective theory has two {dimensionless} couplings $\alpha$ and $\gamma$ determining the holographic and Fermi-liquid-type contributions to the self-energy respectively. We find that irrespective of the choice of the holographic critical sector, there exists a ratio of the effective couplings for which we obtain linear-in-$T$ resistivity for a wide range of temperatures. This scaling persists to arbitrarily low temperatures when $\nu$ approaches unity in which limit we obtain a marginal Fermi liquid with a specific temperature dependence of the self-energy.
\end{abstract}

\maketitle
\section{I. Introduction}
The measurement of the spectral function via ARPES has given us key insights into the nature of elementary constituents in strongly correlated electronic systems which do not admit quasiparticle description, and which also demonstrate a rich variety of  novel superconducting, metallic and insulating phases \cite{RevModPhys.75.473,Vishik_2010,reber2015power}. Phenomenological approaches to model strange metallic behavior have considered spectral functions with the following properties: (i) particle-hole asymmetry, (ii) semilocality (i.e. very mild dependence on the momentum in the direction normal to the Fermi surface), and (iii) nontrivial scaling exponent with the frequency \cite{PhysRevB.78.035103,Lee_2018,RevModPhys.92.031001}. Such features are extremely challenging to obtain from a first principle approach. 

It is quite remarkable that spectral functions with these properties arise in holographic theories at finite density;  the infrared behavior  is described by a Dirac fermion in the near-horizon $AdS_2 \times R^2$ geometry of the black brane dual to the critical sector \cite{Lee:2008xf,Liu:2009dm,Cubrovic:2009ye,Faulkner:2009wj,Faulkner:2010da,Iqbal:2011in} concretely realizing the scenario of \textit{deconfined criticality} \cite{Senthil:2004aza}. Although such theories are essentially gauge theories and the microscopic description is possibly not relevant for material physics, the infrared fixed point with novel scaling behavior at finite density could be universal in a suitable large $N$ limit and thus could provide a first principle realization of emergent non-Fermi liquid behavior.

The holographic approach allows one to dispense with the notion of quasiparticles, which are replaced by the modes of the underlying (emergent) infrared conformal field theory, and this can be very useful to provide a unified picture of transport phenomena \cite{Hartnoll:2016apf}. At the same time, this makes it difficult to understand what are the effective microscopic degrees of freedom of the system. In \cite{PHILLIPS20061634,PhysRevB.77.014512,PhysRevLett.102.056404,PhysRevLett.106.016404,PhysRevB.83.214522,PhysRevB.86.115118} it has been argued that in order to describe the strange metals, it is crucial to consider the effects of intermediate scale physics, especially that of the upper Hubbard band, on the critical sector -- necessitating an approach that is often called \textit{Mottness}. Such effects cannot be reliably modeled by a purely holographic approach.
 Bottom-up models  \cite{Gubser:2009qt,Hartnoll:2009ns,Charmousis:2010zz,Faulkner:2010zz,Lee:2010ii,Davison:2013txa,Zhou:2015dha,Cremonini:2018kla} can holographically engineer a quantum critical sector that can reproduce the scaling of resistivity, Hall angle, etc with temperature but typically only in the low temperature regime (unless further contrived \cite{Jeong:2018tua}).

Based on the key insights provided by Faulkner and Polchinski \cite{Faulkner:2010tq}, a semiholographic effective theory was proposed in \cite{Mukhopadhyay:2013dqa} and further studied in \cite{Doucot:2017bdm}. The proposal of Faulkner and Polchinski was to retain only the infrared part of the holographic sector (i.e. the near horizon geometry) and allow linear hybridization of  some of the bands on the lattice with the bulk holographic fermions; this results in a Fermi surface with low frequency behavior determined by the holographic critical sector, {in particular by the scaling exponent $\nu$ of the holographic fermion}.   In \cite{Mukhopadhyay:2013dqa}, it was shown that if one introduces short-range interactions among the lattice fermions and also more general mutual interactions between the lattice fermions and the holographic fermions with a specific form of large $N$ scaling (see below), the low frequency behavior at the Fermi surface remains unaffected leading to the notion of a \textit{generalized quasiparticle.} The density-density correlations and the case of Coulombic interactions was analyzed in  \cite{Doucot:2017bdm}. In particular, it was found that in presence of a frequency cutoff, the semiholographic theory exhibits well-defined collective excitations within the continuum above a certain momentum threshold and at reasonably low frequencies. It was speculated that these plasmonic excitations may provide a realization of the midinfrared scenario for superconductivity proposed by Leggett \cite{Leggett8365,Leggett:2006aa}.

In this work, we address a key issue in this semiholographic approach by achieving a natural UV completion that interpolates between non-Fermi liquid behavior at low frequencies/temperatures and Fermi liquid behavior at high frequencies/temperatures. The completion does not alter the generalized quasiparticle on the Fermi surface and is also  insensitive to microscopic lattice effects. Such a scenario leads to an effective theory sharing some features with Mottness. Our construction is rather simple and is based on {the assumption} that the two-dimensional band of interest, {in addition to  hybridizing with the infrared critical sector, also} interacts with other bands which have conventional quasiparticle behavior.  We can engineer the interactions in a way that involves only  two effective couplings. {Because the theory is UV complete, observables are well defined without the need of } an {\it ad hoc} frequency cutoff as employed in \cite{Doucot:2017bdm}, however we do reproduce the results of the latter work qualitatively and extend them to finite temperatures. 

{In this letter we use the model described above to compute the dc conductivity at finite temperature. 
Our main result is that we observe} a linear-in-$T$ resistivity for a wide range of temperatures when the scaling dimension $\nu > \sim 0.67$. For $\nu \approx 2/3$, as for instance, this linear scaling regime extends from $0.3 E_F$ to about $20-40 E_F$ when the couplings are small, but when $\nu \approx 1$ it holds at arbitrary low temperatures also. 
Crucially, in our semiholographic approach where we include perturbative degrees of freedom, we are able to obtain the linear-in-$T$ resistivity \textit{irrespectively} of the critical sector (for a wide range of $\nu$) as long as the ratio of the two effective couplings is optimal. Although the range of temperatures where this scaling is valid depends on $\nu$, it is typically very wide.

\section{II. A simple effective theory}
We propose a simple effective theory based on a single band of electrons localized on a two-dimensional plane.  

{The starting point is the model studied  in \cite{Mukhopadhyay:2013dqa,Doucot:2017bdm}. It includes the creation and annihilation operators for the electrons in the band $c^\dagger(\mathbf{k}),(c(\mathbf{k}))$ which are hybridized linearly with a fermionic operator} $\chi_{CFT}(\mathbf{k})$ ($\chi_{CFT}^\dagger(\mathbf{k})$) in the critical sector described by a holographic $AdS_2 \times R^2$ dual geometry. Crucially this band has no direct self-interactions because of a large-$N$ limit discussed in \cite{Mukhopadhyay:2013dqa} in which the self-interactions of the \textit{bath} critical sector [an infrared conformal field theory (IR-CFT)] scale quadratically as $N^2$ while the hybridization coefficient scales linearly with $N$. The backreaction of the metric is suppressed in this limit but the self-energy of the two-dimensional band receives an $\mathcal{O}(N^0)$ correction in the form of a holographic fermionic proagator destroying its quasiparticle nature. It has been shown in \cite{Mukhopadhyay:2013dqa} that the leading low-frequency behavior at the Fermi surface is unaltered as long as we introduce further interactions like $c\chi^3$ which are linear in $c(\mathbf{k})$ and  $c^\dagger(\mathbf{k})$, and scale at most linearly with $N$. In this sense, the semiholographic theory of this band creates a \textit{generalized quasiparticle}. 

One major problem with this version of the semiholographic theory is that the UV behavior is not regular; {one manifestation of this is the fact that} the real part of the density-density correlation function (a.k.a the Lindhard function) {is negative for all frequencies}, unless we impose an artificial frequency cutoff as in \cite{Doucot:2017bdm}. However, we expect the high energy behavior of the theory to be more conventional with the real part of the Lindhard function being positive definite at high frequencies. It is therefore pertinent to look for a modification of this theory in which we obtain {this automatically without implying any specific UV completion.}

Such a modification can be achieved if we assume that the two-dimensional band couples linearly with other bands which preserve their (Landau) quasiparticle characteristics. Denoting the creation (annihilation) operators of one-particle states in these bands as $f_i^\dagger$ ($f_i$) we allow only interactions of the type $cf^3$, $cf^5$, $\cdots$. We thus consider the following Hamiltonian
\begin{eqnarray}\label{Eq:Hamiltonian}
\hat{H} &=&\sum_{\mathbf{k}}\epsilon(\mathbf{k}) c^\dagger(\mathbf{k})c(\mathbf{k})+ N\sum_{\mathbf{k}}\left(g c^\dagger(\mathbf{k})\chi_{CFT}(\mathbf{k})  + c.c.\right) \nonumber\\&&+ N^2 \hat{H}_{IR-CFT} + \sum_{i,j,k}\Big( \lambda_{ijk, \mathbf{k}_1, \mathbf{k}_2, \mathbf{k}_3}\\\nonumber&& c^\dagger(\mathbf{k}_1)  f_i (\mathbf{k}_2)f_j^\dagger(\mathbf{k}_3) f_k (\mathbf{k}_1-\mathbf{k}_2 +\mathbf{k}_3)+ c.c.\Big) + \cdots.
\end{eqnarray}


At leading order in the coupling $\lambda$, the $c$-fermions cannot run in the loops. This leads to factorization of the self-energy of the $c-$band into two parts, namely the contribution from the holographic propagator and a Fermi-liquid type self-energy term {(more details in the Supplemental Material)}. 

In this effective theory the finite-temperature retarded Green's function of the $c-$band thus takes the form
\begin{eqnarray}\label{Eq:Propagator}
G(\omega, \mathbf{k}) &=& \Big(\omega+ i \widetilde{\gamma} (\omega^2 + \pi^2 T^2)+ \widetilde{\alpha}\, \mathcal{G}(\omega)   \nonumber\\ && - \left(\frac{\mathbf{k}^2}{2m} - \frac{k_F^2}{2m}\right)\Big)^{-1}
\end{eqnarray}
with $\widetilde{\gamma} = \mathcal{O}(\lambda^2)$ being the coefficient of the Fermi-liquid type self-energy term, $\widetilde{\alpha}= \mathcal{O}(\vert g \vert^2)$ and  $\mathcal{G}(\omega)$ is the contribution of the holographic sector on the Fermi surface with the form \cite{Iqbal:2009fd,Faulkner:2009wj}
\begin{equation}\label{Eq:Hol-Propagator}
\mathcal{G}(\omega, T) = e^{i(\phi+ \pi \nu/2)} (2\pi T)^\nu \frac{\Gamma(\tfrac12+ \tfrac{\nu}2 - i \tfrac{\omega}{2\pi T})}{\Gamma(\tfrac12 -  \tfrac{\nu}2 - i \tfrac{\omega}{2\pi T})}.
\end{equation}
at finite temperature. Note that the holographic contribution to the self-energy has both real and imaginary parts. Furthermore, $0 < \nu < 1$ is the scaling dimension of the IR-CFT fermionic operator $\chi$ which is related to the mass of the dual bulk fermion. The restriction on $\nu$ is necessary in order for the holographic contribution to be relevant at low frequency. It is easy to check that the spectral function $\rho = - 2 {\rm Im}G$ is non-negative provided $0 < \phi< \pi(1- \nu)$ and $\widetilde\alpha,\widetilde\gamma >0$. 

Note in the high frequency or zero temperature limit $\mathcal{G}(\omega, T)$ is $e^{i\phi}\omega^\nu$ in agreement with the form studied in \cite{Doucot:2017bdm} at zero temperature. In the low energy limit $\mathcal{G}(\omega, T) \approx T^\nu$.  Crucially the limit $\nu \rightarrow 1$ yields a marginal Fermi liquid with a $\omega \log \omega$ term in the self-energy and with a specific type of temperature dependence.

\begin{figure}
   \centering
   
   \includegraphics[width=0.45\textwidth]{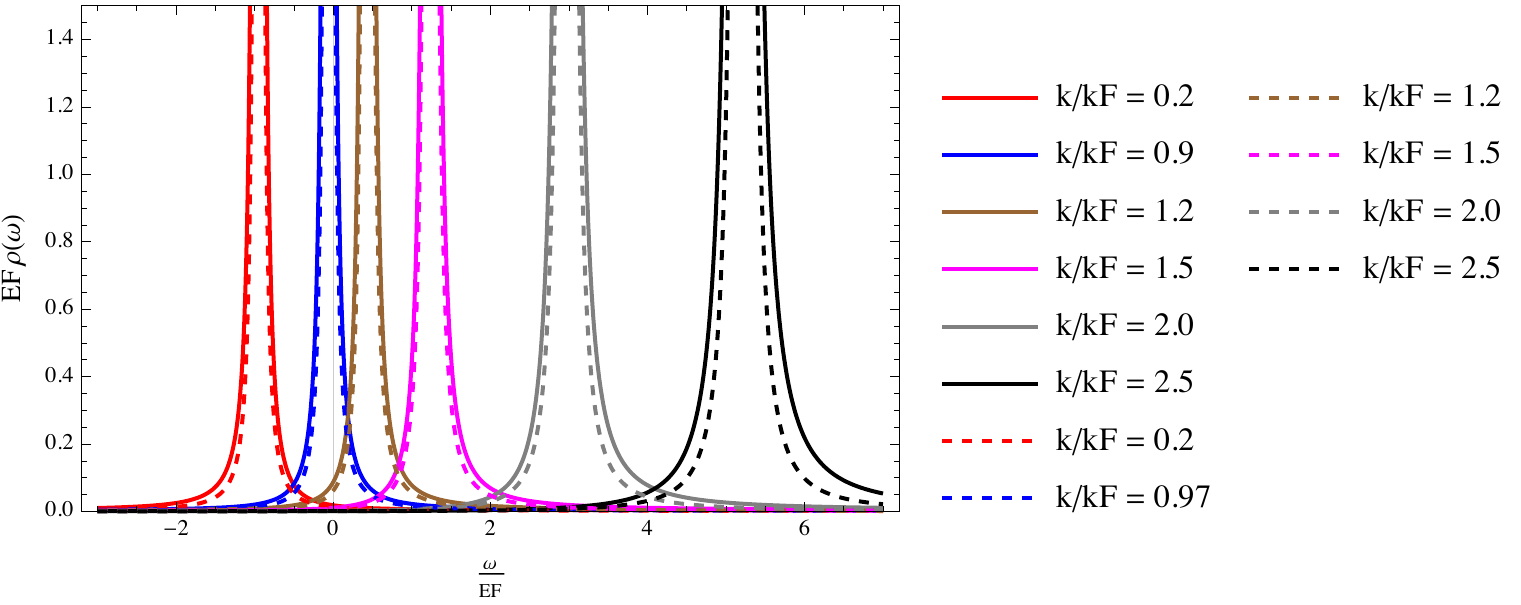}
   \\
   \includegraphics[width=0.4\textwidth]{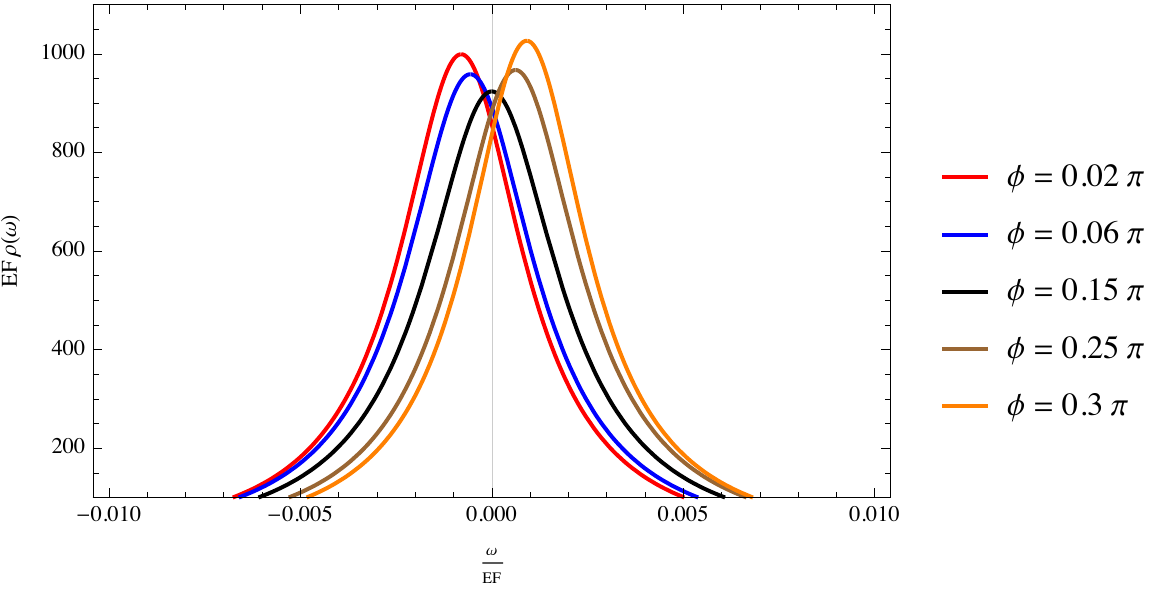}
  \caption{Top: the dimensionless spectral function $E_F \rho(\omega)$ is plotted for various values of $k/k_F$ at $T = 0.5 E_F$, $\nu = 0.7$, $\phi = 0.2\pi$ and $\alpha = 0.016$ with the solid lines corresponding to $\gamma = 0.001$ and the dashed lines corresponding to $\gamma= 0$. Note that the effects of nonvanishing $\gamma$ is significant only for $k$ away from $k_F$. Bottom: here the dimensionless spectral function is plotted for  $T = 0.1 E_F$, $\nu = 0.7$, $\gamma = 0.001$, $\alpha = 0.016$ and various values of $\phi$ between the allowed range $0$ and $(1-\nu)\pi = 0.3 \pi$ (see text) at $k = k_F$. Note that when $\phi \approx 0.15 \pi = \pi(1-\nu)/2$, the spectral function at the Fermi surface is nearly even in $\omega$. }\label{Fig:Spectral}
\end{figure}

{The model contains a single intrinsic scale, given by the Fermi energy $E_F$. } 
It is useful to rewrite the propagator \eqref{Eq:Propagator} in terms of dimensionless variables $x := \omega /E_F$ and $\mathbf{y} := \mathbf{k}/k_F$ in the form
\begin{eqnarray}
G(x, \mathbf{y}) &=& E_F^{-1}\Big( x + i \gamma (x^2 + (\pi x_T)^2) + \alpha e^{i(\phi+ \pi \nu/2)}\nonumber\\&& (2\pi x_T)^\nu \frac{\Gamma(\tfrac12+ \tfrac{\nu}2 - i \tfrac{x}{2\pi x_T})}{\Gamma(\tfrac12 -  \tfrac{\nu}2 - i \tfrac{x}{2\pi x_T})}- \left(\mathbf{y}^2 - 1\Big)\right)^{-1}
\end{eqnarray}
with $x_T := T/E_F$, $\alpha = \widetilde{\alpha} E_F^{-(1-\nu)}$ and $\gamma = \widetilde\gamma E_F$. Note $\alpha$ and $\gamma$ are dimensionless. Some plots of the spectral function are shown in Fig. \ref{Fig:Spectral}.
 
{The density-density correlation functions (the Lindhard function $\mathcal{L}(q, \Omega)$) can be readily computed from the above spectral function at finite temperature. We reproduce the qualitative features of the zero temperature density-density correlations computed earlier in \cite{Doucot:2017bdm}. Even at finite temperatures $T\approx 0.5 E_F$, the edges of the continuum are still prominent when $\alpha, \gamma \ll 1$ although the response has sufficient support outside the continuum. Furthermore, there exist well-defined plasmonic excitations for $q < 2.5 k_F$ which have support inside the continuum when $\approx 2 k_F < q < \approx 2.5 k_F$. The dispersion relation of the plasmonic modes is approximately linear. (See Supplemental Material.)} Furthermore, as noted in \cite{Doucot:2017bdm}, it is important to consider $1/2 < \nu \leq 1$ to avoid ultraviolet issues that destabilizes the infrared effective theory.

Finally we emphasize that it is necessary for $\gamma$ to be sufficiently small and also $\alpha < 1$ for the effective theory to be applicable at high frequencies and temperatures. If this is not the case, the effects of $c-$fermion running in loops will spoil the separation of the self-energy into holographic and Fermi-liquid terms. {Since both of these effective couplings are irrelevant on the Fermi surface \cite{Mukhopadhyay:2013dqa}, our assumption is consistent.}

 \section{III. DC conductivity}
 The dc conductivity can be computed readily from the simple formula
 \begin{eqnarray}
  \sigma_{dc} \approx \frac{e^2}{2\hbar}\int \frac{d\omega}{2\pi} \int \frac{d^2k}{4\pi^2} \,\, k^2\rho(\omega,\mathbf{k},T)^2\left(- \dfrac{\partial n_F(\omega,T)}{\partial \omega}\right).
 \end{eqnarray}
 Above $n_F(\omega, T)$ denotes the Fermi-Dirac distribution function. Here we have used the exact bare vertices for the coupling with the photon and not simply its value on the Fermi surface which give the $k^2$ term in the integrand. {We do not impose any cutoff in the integrations.}
 
 \begin{figure}
   \centering
   
   \includegraphics[width=0.51\textwidth]{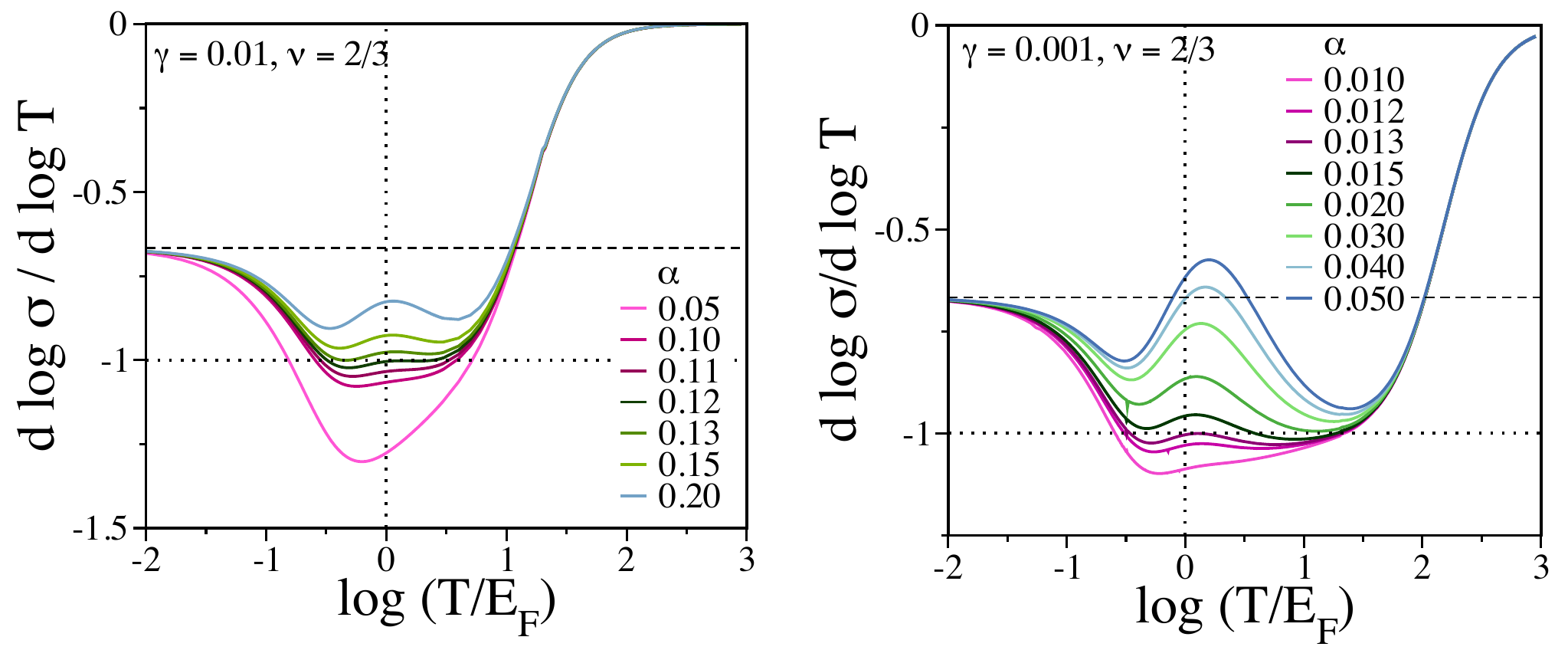}
     
\caption{The temperature dependence of the scaling of the conductivity with the temperature for $\nu = 2/3$ is shown for various values of $\alpha$ when $\gamma = 0.01$ (on left) and $\gamma = 0.001$ (on right). The upper horizontal dashed lines in both plots indicate that the scaling  at very low $T$ is $-2/3$. The linear-in-$T$ scaling of resistivity appears in midtemperature regime when $\alpha \approx 13 \gamma$. 
}\label{Fig:Nu2by3}
\end{figure}
  
It is quite easy to see that the low temperature behavior of the conductivity should be determined by the holographic critical sector so that $\sigma_{dc}(T) \approx T^{-\nu}$ for $T \ll 0.1 E_F$. For $T > 0.1 E_F$, the Fermi liquid contribution to the self-energy is influential when $0.001 \leq \gamma \leq 0.01$. {At high temperatures $T \gg E_F$, the dc conductivity becomes almost independent of the holographic critical sector and therefore does not depend on $\alpha$, $\nu$ and $\phi$.  In this regime the scaling exponent decays rapidly implying that the rate of decay of the dc conductivity with temperatures slows down. We find that a scaling regime where $\sigma_{dc} \approx T^{- \widetilde{\nu}}$ with $\widetilde{\nu}\neq \nu$ and approximately independent of the temperature emerges for $\nu > \sim 0.67$ to a very good approximation at intermediate temperatures. Interestingly, $\widetilde{\nu} \approx 1$ implying linear-in-$T$ resistivity in this midtemperature regime which stretches to arbitrary small values of temperatures when $\nu$ approaches $1$ where we obtain a marginal Fermi liquid with a specific temperature-dependent self-energy. (Note we restrict ourselves to $1/2 < \nu < 1$ for reasons mentioned above.)}

{Furthermore, we also find that the best approximation to the scaling in midtemperature regime is obtained when the phase $\phi$ in Eq. \eqref{Eq:Propagator} takes its value around $\pi(1-\nu)/2$ (the midpoint of the allowed range of values). We therefore choose $\phi$ to be around this value for the rest of this paper. We observe that in this case the spectral function at the Fermi surface is approximately even in $\omega$ as shown in Fig. \ref{Fig:Spectral}.}
  
 \paragraph{\textbf{Case 1 ($\boldsymbol{0.5 < \nu < \sim 0.8}$):}} 
  It is useful to first study the representative case of $\nu = 2/3$ plotted in Fig. \ref{Fig:Nu2by3} where the independence of the high temperature behavior on the holographic sector is manifest. In the intermediate temperature regime, the scaling exponent ${\rm d}\log \sigma_{dc}/{\rm d}\log T$ is nonmonotonic. However, crucially we observe that when $\alpha \approx 13\gamma$, there is a scaling regime in which the scaling exponent is temperature independent for both $\gamma = 0.01$ and $\gamma = 0.001$ and furthermore it is approximately $-1$ implying linear-in-$T$ resistivity. Remarkably, the range of temperatures where this occurs is quite wide extending from about $0.3 E_F$ to nearly $10 E_F$ when $\gamma  =0.01$ and nearly $20 E_F$ when $\gamma  =0.001$. 

 \begin{figure}
   \centering
  \includegraphics[width=0.45\textwidth]{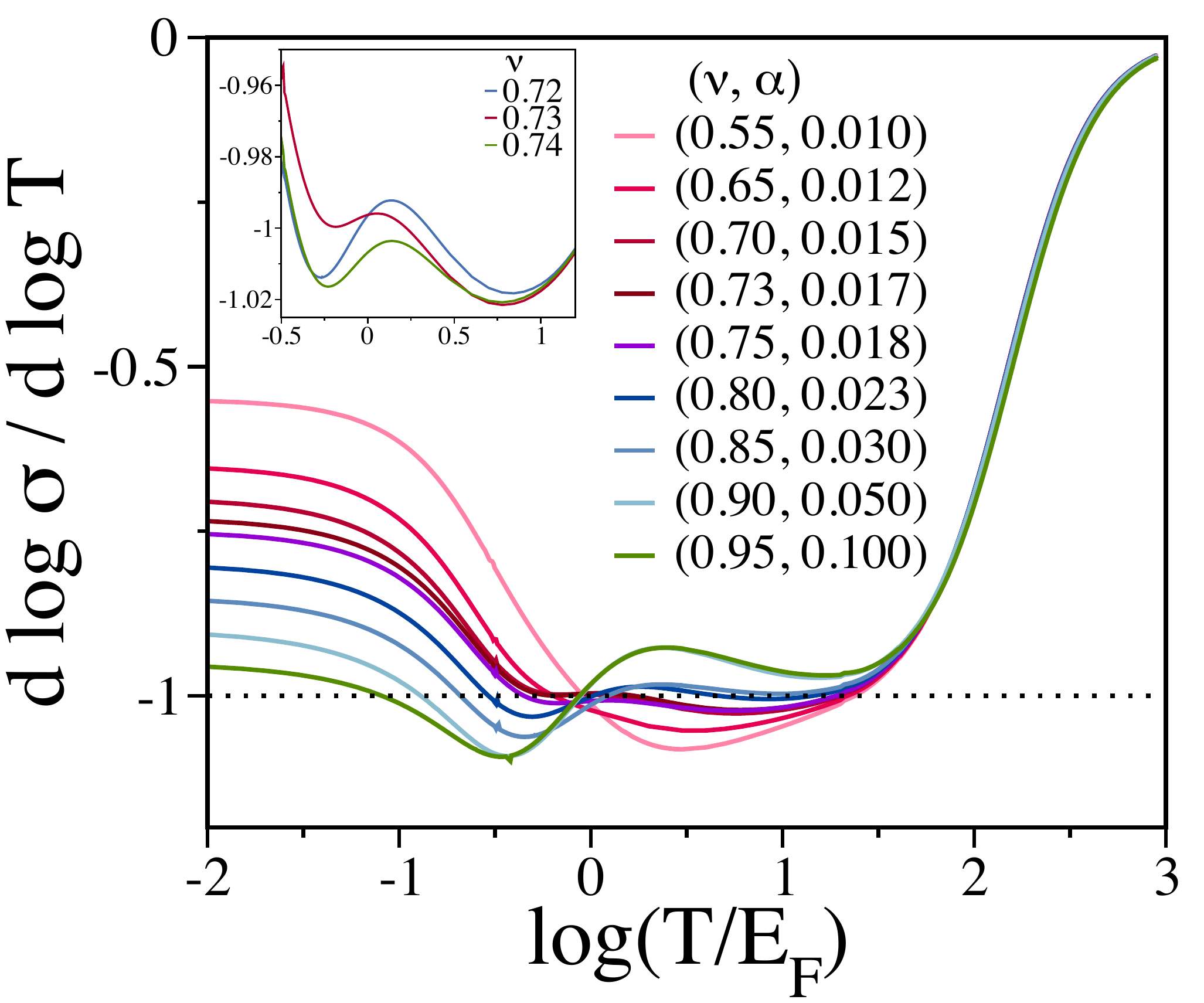}

\caption{The scaling exponent of the dc conductivity is plotted for various values of $\nu$ between $0.5$ and $1$ at $\gamma = 0.001$. Above $\alpha$ has been optimally fine-tuned. The inset plot shows that the best linear-in-$T$ scaling is obtained for $\nu = 0.73, \alpha = 0.017$. 
}\label{Fig:NuVarious}
\end{figure}
\begin{figure}
   \centering
     \includegraphics[width = 0.35\textwidth]{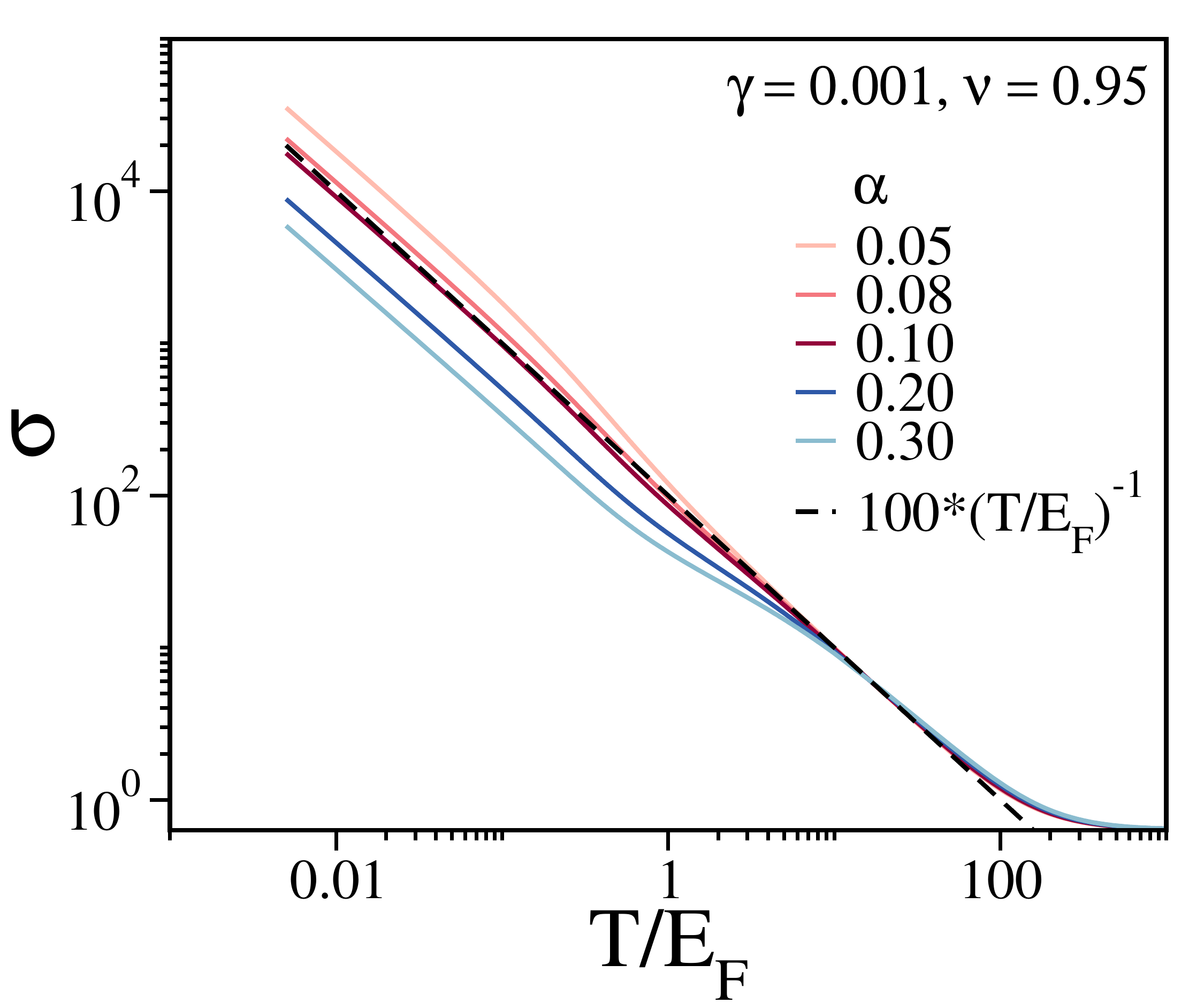}
\caption{The temperature dependence of the dc conductivity is plotted above for $\nu = 0.95$, $\gamma = 0.001$ and various values of $\alpha$. The best approximation to linear-in-$T$ resistivity is obtained for $\alpha = 0.1$. The scaling exponent for this case has been plotted as a function of the temperature in Fig \ref{Fig:NuVarious}.}\label{Fig:Nu0p95}
\end{figure}
Such a scaling regime at intermediate temperature regime does not arise for $\nu < \sim0.67$. In Fig. \ref{Fig:NuVarious}, we report the dependence of the scaling exponent on the temperature at various values of $\nu$ ranging between $0.5$ and $1$ for $\gamma =0.001$ and for those values of $\alpha$ where we get the best approximation to linear scaling of the dc resistivity with the temperature. Clearly, for $\nu = 0.55$, there is no scaling regime for $T > 0.1 E_F$. The best approximation to the linear-in-$T$ resisitivity in the temperature range that includes $E_F$ is obtained when $\nu = 0.73$ and $\alpha = 0.017$  (see inset plot of Fig \ref{Fig:NuVarious} where it shows that the scaling exponent varies between $0.98$ and $1.02$ for $0.3 E_F < T < 30 E_F$).

 Thus we obtain linear-in-$T$ resistivity to a remarkable approximation in the midtemperature regime $T_{min} < T < T_{max}$ with $E_F$ included in this range for $\sim 0.67 < \nu < \sim 0.8$ if we choose the optimal ratio $\alpha/\gamma$ for each $\nu$ while setting $\phi $ near $\pi(1-\nu)/2$ as mentioned above.{The higher end of this scaling regime ($T_{max}$) strongly depends on $\gamma$ but the lower end ($T_{min}$) depends only mildly on $\gamma$ and the parameters of the holographic sector.} 



 \paragraph{\textbf{Case 2 ($ \sim 0.8 < \nu \leq 1$):}} For higher values of $\nu$ we can still obtain the linear-in-$T$ resistivity in the  midtemperature regime including $ T \approx E_F$ by tuning $\alpha/\gamma$ for each choice of $\nu$ as shown in Fig. \ref{Fig:NuVarious}. However, the scaling is less accurate, i.e. $-1 \pm 0.1$ percent instead of $-1 \pm 0.05$. In the case of $\nu = 0.85$, as for instance, we get an excellent linear-in-$T$ resistivity only at higher range of temperatures between $3E_F$ and $30E_F$. However, if we allow for $10$ percent variation of the scaling exponent, then the scaling regime stretches to arbitrarily low temperatures as illustrated in the case of $\nu = 0.95$ in Fig. \ref{Fig:Nu0p95}. This continues to hold as we approach the marginal Fermi liquid $\nu \approx 1$. 
 
{Thus our effective field theory approach that is justified by the Wilsonian renormalization group shows that we can achieve the linear-in-$T$ resistivity  for $\nu> \sim 0.67$ at intermediate temperatures just by fine tuning the ratio of the two effective dimensionless couplings.}

\section{Discussion}
{Our effective semiholographic approach, that shares common features with Mottness, produces the linear-in-$T$ resistivity over a very wide range of temperatures irrespectively of the holographic critical sector provided $\nu > \sim 0.67$ when we tune the ratio of the two couplings optimally. Although the Fermi energy provides the unique energy scale of our model, in order to match it with a material we should probably use a lower scale, {eg. where the self-energy becomes strongly $k$-dependent.}
In any case, the model is just the first step to a more viable theory applicable to real-world strange metals.
}


{It is useful to compare our approach to some recently discussed models involving a lattice of (complex) SYK quantum dots exchanging fermions via hopping  \cite{PhysRevLett.119.216601,PhysRevX.8.021049,PhysRevX.8.031024,PhysRevLett.121.187001,Cha_2020,PhysRevLett.122.186601,PhysRevLett.123.066601} which also can reproduce linear-in-$T$ resistivity (see also \cite{PhysRevResearch.2.033434}). A heuristic connection with our approach readily emerges from the observation that (nearly) $AdS_2$ holography can capture many aspects of SYK systems \cite{Sarosi:2017ykf}. Thus a lattice of $AdS_2$ throats representing a fragmented $AdS_2 \times R^2$ geometry recently proposed as a model for quantum black hole microstates \cite{Kibe:2020gkx} could be actually also relevant for our approach. {Aided via DMFT methods in which the Anderson impurity atom is replaced by a single $AdS_2$ throat {(without any compact/noncompact tensor part)}, we aim to understand the thermodynamic reason for why a certain ratio of the couplings can be preferred. We also plan to explore magneto-transport and the superconducting instability.}

 \begin{acknowledgments}
It is a pleasure to thank Johanna Erdmenger, Rene Meyer, Ronny Thomale, Mukul Laad and Shantanu Mukherjee  for helpful discussions. A.M. acknowledges support from the Ramanujan Fellowship and ECR award of the Department of Science
and Technology of India. A.M. and G.P. also acknowledge generous support from IFCPAR/CEFIPRA funded project no 6403. 
\end{acknowledgments}

 \onecolumngrid
\appendix
\section{Derivation of the fermionic propagator}
The action of our effective theory is simply
\begin{eqnarray}
S &=& \int {\rm d} t\int{\rm d}^2 k \,c^\dagger({\mathbf{k}})(i\partial_t - \epsilon_{\mathbf{k}})c({\mathbf{k}}) \nonumber\\&&
+N \int {\rm d} t\int{\rm d}^2 k \left(  g c^\dagger({\mathbf{k}})\chi_{\rm CFT}({\mathbf{k}})+ c.c.\right) + N^2 S_{\rm hol}\nonumber\\&&+ \int {\rm d} t\int{\rm d}^2 k_1\int{\rm d}^2 k_2\int{\rm d}^2 k_3 \,\lambda \,c^\dagger({\mathbf{k}_1})f({\mathbf{k}_2})f^\dagger({\mathbf{k}_3})f({\mathbf{k}_3 + \mathbf{k}_1- \mathbf{k}_2})\nonumber\\&& + h.c.
\end{eqnarray}
with $\epsilon_{\mathbf{k}} = (k^2 - k_F^2)/2m$ and where we have suppressed the $k$-dependence of $\lambda$. In the large $N$ limit, the self-energy correction to the $c$-fermion is
\begin{eqnarray}
\Sigma_c(\mathbf{k},\omega) &=& N^2 \vert g\vert^2 \frac{1}{N^2}\mathcal{G}(\omega,\mathbf{k}) +\,\, {\rm  loop\,\, diagrams \,\,with\,\, c\,\, and \,\,f\,\, fermions} + \,\,\mathcal{O}(N^{-2})\nonumber\\
 &=& \vert g\vert^2 \mathcal{G}(\omega,\mathbf{k}) +\,\, {\rm  loop\,\, diagrams \,\,with\,\, c\,\, and \,\,f\,\, fermions} + \,\,\mathcal{O}(N^{-2}).
\end{eqnarray}
Above $\mathcal{G}(\omega,\mathbf{k})$ is the temperature dependent two-point function of $\chi_{CFT}$ which is obtained from solving the Dirac equation of the dual holographic fermion in the bulk geometry. The above follows from the following features: (i) loops in the holographic bulk geometry are suppressed in the large $N$ limit, (ii) tree diagrams in the bulk with $n$-points at the boundary scale as $N^{2 -n}$ implying suppression of bulk vertices also. Crucially in the large $N$ limit, the self-energy contribution is simply the sum of the holographic contribution and a Fermi liquid type term arising from loops with the $\lambda$-vertices. Therefore, we obtain that
\begin{eqnarray}
\Sigma_c(\mathbf{k},\omega) &=& \vert g\vert^2 \mathcal{G}(\omega,\mathbf{k}) + \lambda^2 b (\omega^2 + \pi^2 T^2) + \,\,\mathcal{O}(N^{-2}, \lambda^2\vert g\vert^2, \lambda^4).
\end{eqnarray}
Furthermore, if the dual holographic geometry is of the type $AdS_2 \times R^2$ (note backreaction is suppressed in the large $N$ limit and by the subscript CFT in $\chi_{CFT}$ we imply a scale invariant infrared fixed point and not a $2+1-$dimensional CFT), then the $k$-dependence arises from $\nu(k)$, the $k$-dependent scaling dimension of $\chi_{CFT}$ which is determined by the  $k$-dependent $AdS_2$ mass of the dual bulk fermion. Near the Fermi surface, $\nu(k)$ has a weak dependence on $k$, and therefore we can assume $\nu(k)\approx\nu(k_F)$. Thus we obtain the fermion propagator given by Eq. (2) in the main text. 

We note that one may also think of the $AdS_2$ geometry as merely providing a holographic description of $SYK$-type quantum dots smeared over the lattice. Assuming that the mutual interactions between these quantum dots is suppressed, while a suitable operator of the holographic quantum dot sector, namely $\chi_{CFT}$ couples to the $c$-fermion, the holographic contribution to the self-energy is purely local (in space) and is thus $k$-independent in the large $N$ limit. 

It is worth noting that if we diagonalize the quadratic terms  of the action (see Supplemental Material of Ref [31]), then we obtain two propagating fermion modes of which only one has a Fermi surface singularity and whose overlap with  $\chi_{CFT}$ is suppressed by $\mathcal{O}(N^{-1})$, while the other one vanishes on the Fermi surface like the $AdS_2$ propagator $\mathcal{G}(\omega) \approx \omega^\nu$ at $T=0$. Note coincidence of poles and zeroes (more generally branch points) on Fermi surface is a feature of Mottness. 

\section{The generalized Lindhard function and the plasmonic excitations}
The density-density response function $\mathcal{L}$ (generalized Lindhard function) is schematically
\begin{equation}
\mathcal{L}(\Omega, q)= -2 i \int_k \int_\omega G(\omega_+, \mathbf{k}_+) G(\omega_-, \mathbf{k}_-),
\end{equation}
where  $\omega_{\pm} = \omega \pm \Omega/2$ and $\mathbf{k}_\pm = \mathbf{k} \pm q/2$. In terms of the variables introduced in the main text, the imaginary part of the Lindhard function explicitly is
\begin{eqnarray*}
\text{Im} \mathcal{L} (x_\Omega, y_q , x_T)& =& 4 \int_{-\infty}^\infty dx \int d^2 y\,  \text{Im}G \left( x+ \tfrac{x_\Omega}2, y + \tfrac{y_q}2, x_T\right) \text{Im}G \left( x-\tfrac{x_\Omega}2, y - \tfrac{y_q}2, x_T\right) \left(n_F (x+ \tfrac{x_\Omega}2, x_T)  - n_F (x - \tfrac{x_\Omega}2, x_T) \right)
\end{eqnarray*}
The plot of Im$\mathcal{L}$ as a function of $x_\Omega$ is shown in Fig.~\ref{ImL}.
 \begin{figure}[ht]
\centering
\includegraphics[width=1.0\textwidth]{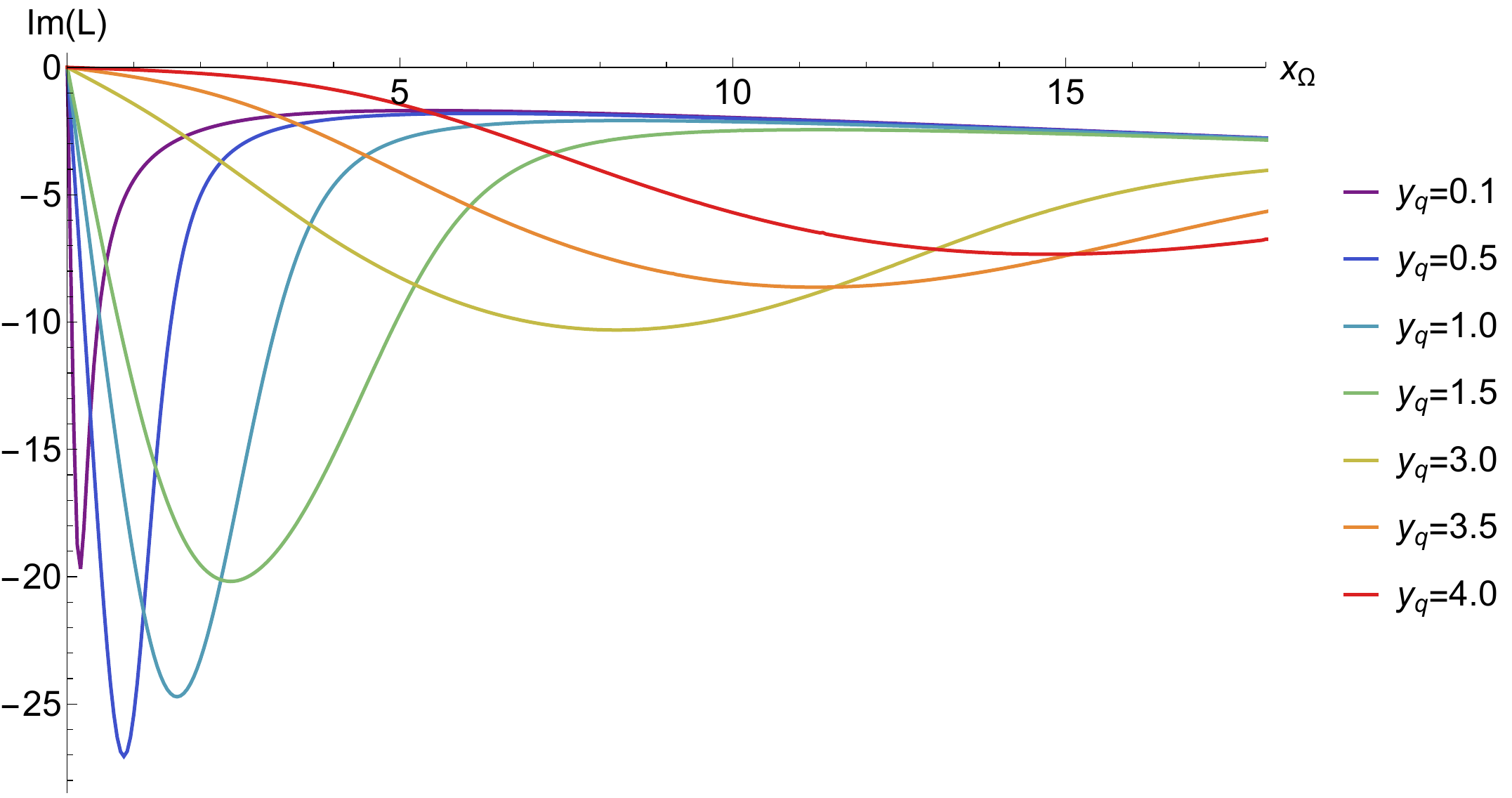}
\caption{Imaginary part of the Lindhard function as a function of $x_\Omega$. We have set $x_T=0.5$, $\alpha= 0.12$ and $\gamma=0.01$.}\label{ImL}
\end{figure}
The continuum extends between $0 < \Omega < q^2/2m + 2 q k_F/m $, i.e. between $0< x_\Omega <y_q^2 + 2 y_q$ for $q \leq 2k_F$ (i.e. $y_q \leq 2$) and between $ q^2/2m - 2 q k_F/m< \Omega < q^2/2m + 2 q k_F/m $, i.e. between $y_q^2 - 2 y_q< x_\Omega <y_q^2 + 2 y_q$ for $q > 2k_F$ (i.e. $y_q > 2$). As for instance, for $y_q = 1$, the maximum support is in the region $0 < x_\Omega < 3$ and for $y_q = 3$ the range of maximum support is $3 < x_\Omega < 15$. It is clear from Fig.~\ref{ImL} that Im$\mathcal{L}$ is supported maximally within the continuum although the edges are blurred out especially for higher values of $q$.

We compute the real part of the Lindhard function using the Kramers-Kr\"{o}nig relation:
\begin{align}
\text{Re}\mathcal{L} (x_\Omega) = \lim_{\epsilon\to0}\frac1{\pi} \int_0^\infty dx\, \text{Im}\mathcal{L}(x) \left(\frac{ x - x_\Omega}{(x - x_\Omega)^2 + \epsilon^2}  +  \frac{ x +x_\Omega}{(x +x_\Omega)^2 + \epsilon^2} \right)\, .
\end{align}
 We put a cut-off on the $x-$integral at $x=30$ and set $\epsilon = 10^{-4}$ while evaluating the above integral numerically. The plot of Re$\mathcal{L}$ is shown in Fig.~\ref{ReL}.
 \begin{figure}[h]
\centering
\includegraphics[width=1.0\textwidth]{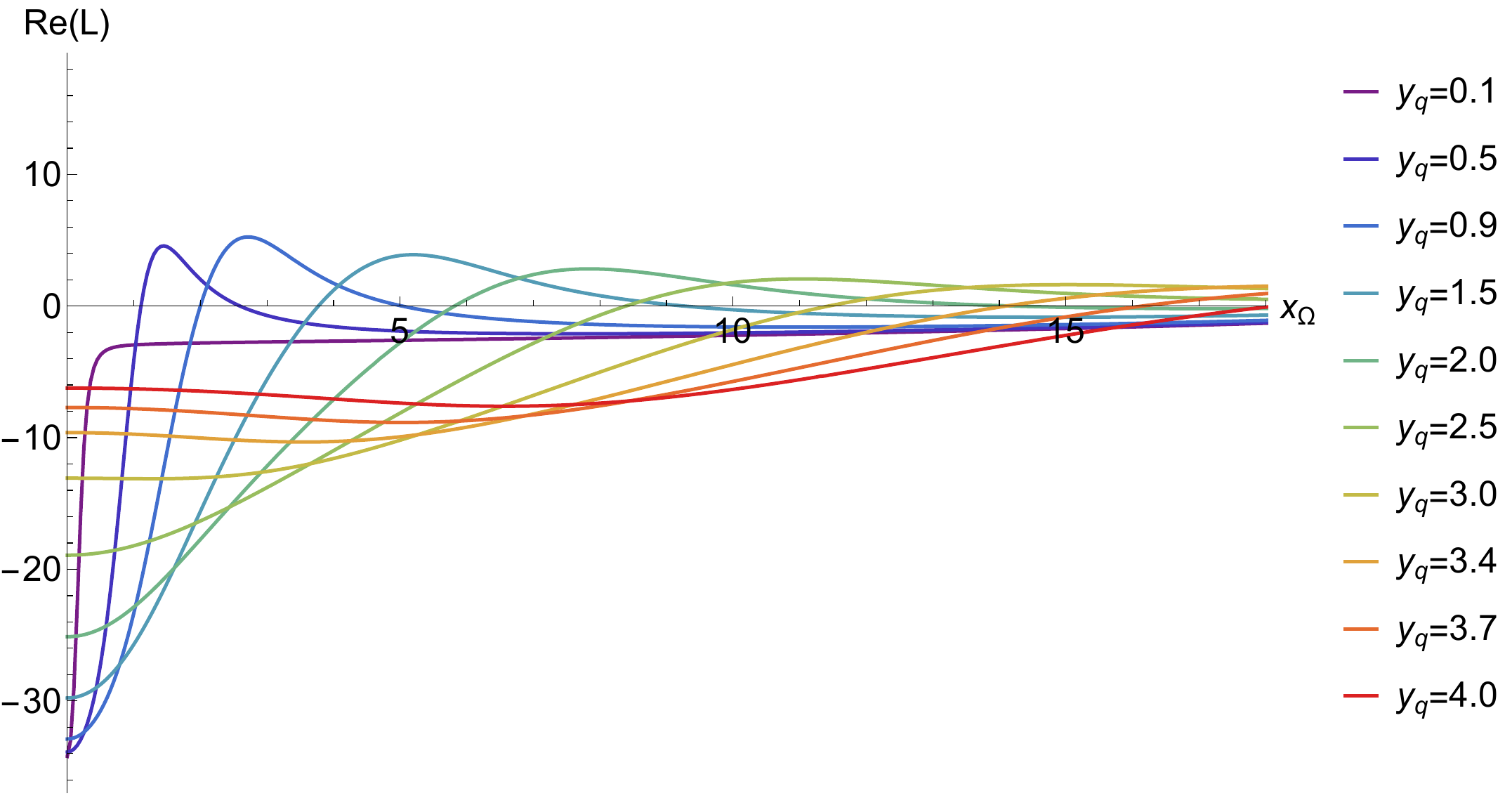}
\caption{Real part of the Lindhard function as a function of $x_\Omega$. We have set $x_T=0.5$, $\alpha= 0.12$ and $\gamma=0.01$.}\label{ReL}
\end{figure}
 
 The bubble resummation produces the improved generalized Lindhard function which is given by
 \begin{align}
 \mathcal{L}^{\text{imp}} (x_\Omega, y_q, x_T)= \frac{\mathcal{L}(x_\Omega, y_q, x_T)}{1- V(y_q) \mathcal{L}(x_\Omega, y_q, x_T)},
 \end{align}
 where $V(y_q)$ is the Coulomb potential. In 2D, this is inversely proportional to $y_q$. The real and imaginary parts of the improved Lindhard function are shown in Fig.~\ref{Limp}.
\begin{figure}[h]
\centering
\includegraphics[width=1.0\textwidth]{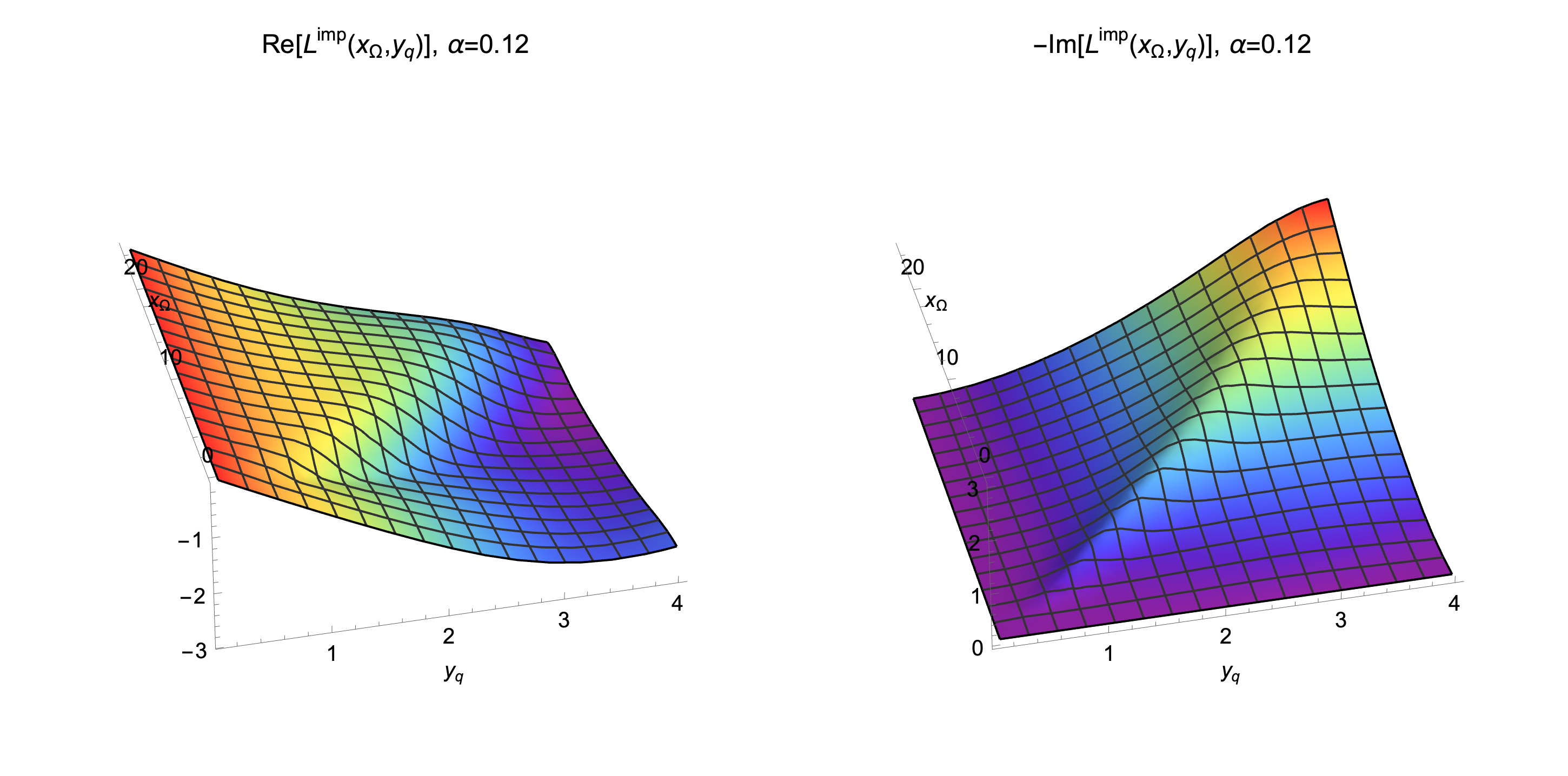}
\caption{Real  and imaginary part of the improved Lindhard function as a function of $x_\Omega$ and $y_q$ when $x_T =0.5$. We have set $\alpha= 0.12$ and $\gamma=0.01$.}\label{Limp}
\end{figure}
Fig.~\ref{2dImLimp} shows the contour plot for imaginary part of the improved Lindhard function. In these figures we have chosen the values of the couplings and the temperature such that they lie in the regime of linear-in-$T$ resistivity. The well defined plasmonic poles with linear dispersion relation are prominently visible. Around $q=2k_F$, the plasmonic excitations have sufficient support inside the continuum which could not have occurred in the case of a Fermi liquid. 
\begin{figure}[h]
\centering
\includegraphics[width=0.6\textwidth]{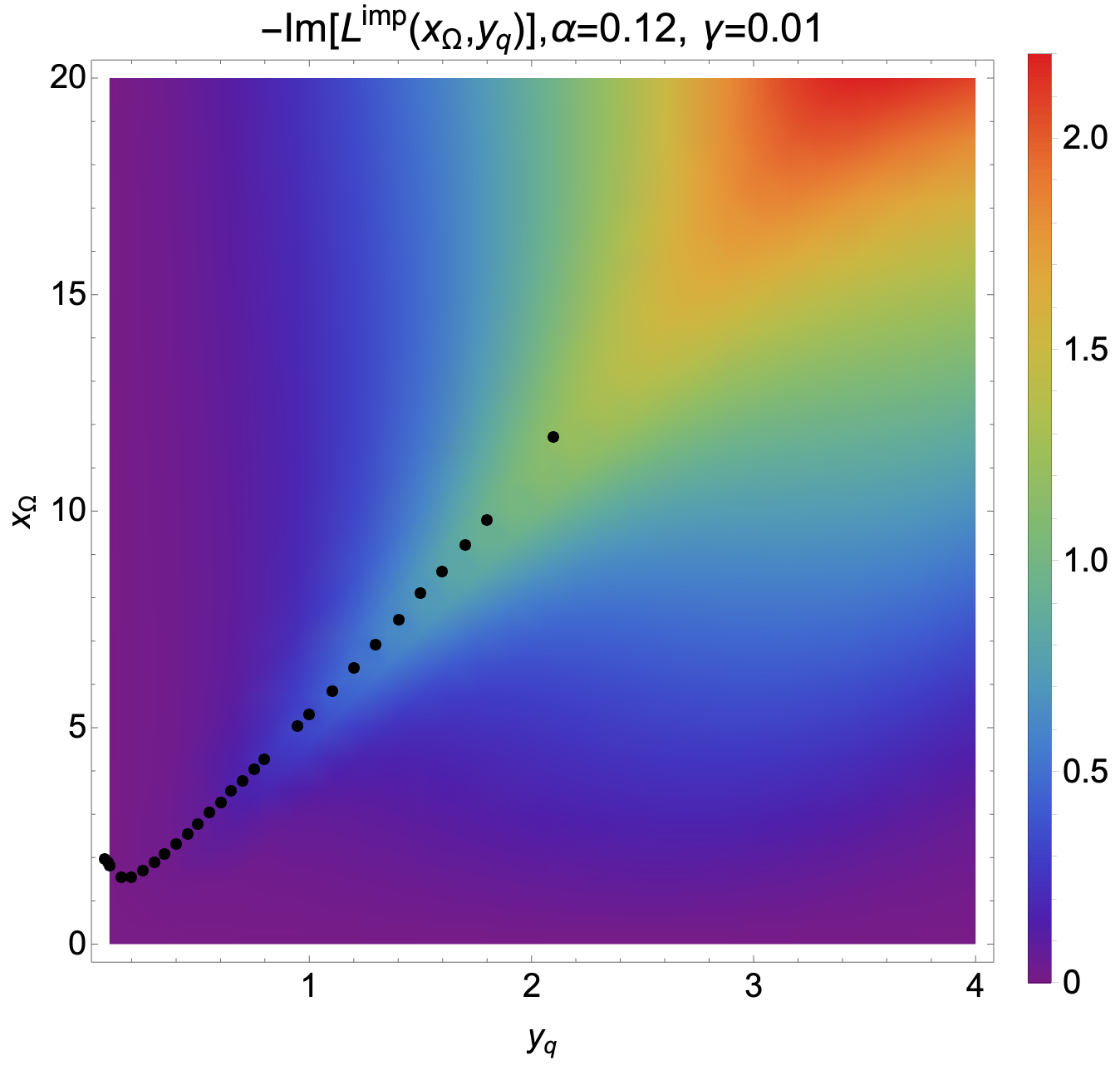}
\caption{Density plot of imaginary part of the improved Lindhard function with the maxima shown as the black dots. We have set $x_T=0.5$, $\alpha= 0.12$ and $\gamma=0.01$.}\label{2dImLimp}
\end{figure}
\clearpage
\bibliography{cond-refs}

\end{document}